\documentclass[dvips,twoside,fleqn]{article}
\usepackage{times}
\usepackage{amstext}
\usepackage{amssymb}
\usepackage{espcrc2}
\usepackage{graphicx}
\title{Topology in CP$^{N-1}$ models: a critical comparison of 
different cooling techniques}

\author{B.~All\'es\address{Dipartimento di Fisica, 
Universit\`a di Milano--Bicocca and
INFN--Milano, I--20133 Milano, Italy},
L.~Cosmai\address{INFN-Bari, I--70126 Bari, Italy},
M.~D'Elia\address{Dipartimento di Fisica, Universit\`a
di Pisa and INFN--Pisa, I--56127 Pisa, Italy},
A.~Papa\address{Dipartimento di Fisica, Universit\`a della Calabria and 
INFN--Cosenza, I--87036 Arcavacata di Rende, Cosenza, Italy}}

\begin{document}
\pagestyle{empty}

\begin{abstract}
Various cooling methods, including a recently introduced one
which smoothes out only quantum fluctuations larger than a given
threshold, are applied to the study of topology in 2d CP$^{N-1}$ models. 
A critical comparison of their properties is performed.
\end{abstract}

\maketitle

Two-dimensional CP$^{N-1}$ models play an important role in quantum field 
theory because they share many properties with QCD. In particular,
they possess instanton classical solutions and one can define 
topological charge and susceptibility in analogy with QCD. 
CP$^{N-1}$ models represent therefore 
a useful theoretical laboratory to investigate numerical methods to be 
eventually applied to study the topology in QCD. One of the most powerful tools 
for the study of the topological structure of the vacuum is the ``cooling'' method.
It consists in measuring the topologically relevant quantities on the ensemble
of lattice configurations obtained by replacing each equilibrium configuration
by the one resulting after a sequence of local minimizations of the action. 

The aim of this work is to get insight into a new cooling method 
first adopted in Ref.~\cite{PPS99}, by comparing it 
with the ``standard''~\cite{Tep86} and with its 
``controlled'' version adopted by the Pisa group~\cite{CDPV89}.

We chose the standard discretization for the action of the 2d CP$^{N-1}$ model:
\begin{equation}
S^L \!= \!-N\beta\sum_{n,\mu}\left(\bar z_{n+\mu}z_n\lambda_{n,\mu} \!+\!
   \bar z_nz_{n+\mu}\bar\lambda_{n,\mu} \!-\! 2\right)
\end{equation}
where $z_n$ is an $N-$component complex scalar field with $\bar z_n \cdot z_n = 1$, 
$\lambda_{n,\mu}$ is a U(1) gauge field satisfying 
$\bar\lambda_{n,\mu}\lambda_{n,\mu} = 1$ and $\beta=1/(N g)$, with 
$g$ the lattice coupling. 
We used the standard action both in Monte Carlo simulations and in the cooling
instead of any improved lattice action, since we wanted to test the cooling 
techniques in a situation where cutoff effects are large.
The lattice topological charge density was defined as 
\begin{equation}
q^L(x) \!= \!-{i\over 2\pi}\sum_{\mu\nu} \epsilon_{\mu\nu}
   {\rm Tr}\left[ P(x)\Delta_\mu P(x)
   \Delta_\nu P(x) \right],
\end{equation}
with
\begin{eqnarray}
\Delta_\mu P(x) &\equiv& \frac{P(x{+}\mu) - P(x{-}\mu)}{2}~, \nonumber
\\
P_{ij}(x) &\equiv& \bar z_i(x) z_j(x) \,,
\end{eqnarray}
giving for the lattice topological susceptibility
\begin{equation}
\chi^L \equiv \biggl\langle \sum_x q^L(x)q^L(0) \biggr \rangle =
{1\over L^2} \left< \left(Q^L\right)^2 \right>\;,
\end{equation}
where $Q^L\equiv\sum_xq^L(x)$ and $L$ is the lattice size.
The cooling algorithm consists in assigning to each lattice variable
$z_n$ and $\lambda_{n,\mu}$ a new value $z_n^{\mathrm{new}}$ and 
$\lambda_{n,\mu}^{\mathrm{new}}$ which locally minimizes the action, keeping all 
other variables fixed. In the ``standard cooling'' these
replacements are unconstrained. We will call ``new cooling'' the one for 
which the replacements are done only if the angle $\alpha$ between the new and the old 
field variables, is larger than a given value $\delta$, and 
``Pisa cooling'' the one for which the local minimization is performed
with the constraint $\alpha \le \delta_{\rm Pisa}$.
Notice that between the ``Pisa cooling'' and the ``new cooling'' there is a 
substantial difference: while ``Pisa cooling'' acts first on the smoother 
fluctuations, the ``new cooling'' performs local minimizations only
if these fluctuations are larger than a given threshold. Moreover
the ``new cooling'' automatically stops when there are no more
fluctuations beyond the threshold. It should be pointed out that any cooling 
procedure causes a partial loss of the topological content of the cooled 
configuration (namely ``small instantons''). However this loss occurs at a fixed 
scale in lattice units and thus vanishes in the continuum limit, unless the 
instantons distribution is ultraviolet singular (as in the case of CP$^{1}$).

\begin{figure}[t]
\begin{center}
\includegraphics[width=0.45\textwidth]{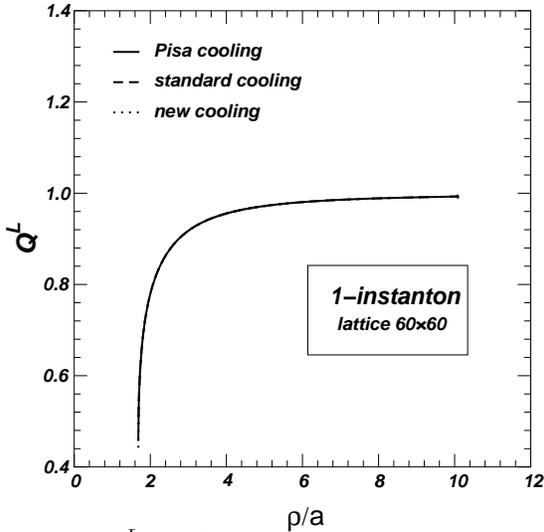}
\end{center}
\vspace{-1.54truecm}
\caption{$Q^L$ {\it vs} $\rho/a$ under the three different cooling procedures.
The instanton size is determined by $S^{L\;{\mathrm{max}}} =
2/(\rho/a)^2$, ($a$ is the lattice spacing).The three curves fall exactly 
on top of each other.} 
\label{q_rho}
\end{figure}

We considered first an ``artificial'' 1-instanton configuration discretized 
on the lattice. Although the three different cooling procedures 
act differently, since they start to deform the configuration in different
regions, the curves for $Q^L$ and $S^L$ under the three coolings 
as a function of the instanton size $\rho$ fall on top of each other (see 
Fig.~\ref{q_rho}).
We determined the topological susceptibility using the ``new cooling'' 
and compared the results with those from 
the ``field theoretical method''~\cite{CDPV90}. In the field theoretical approach 
one has 
\begin{eqnarray}
\chi^L(\beta) &=& a^2 Z(\beta)^2\chi^{\mathrm{cont}} + M(\beta) + O(a^4) \nonumber \\ 
M(\beta) &=& a^2 \,A(\beta) \langle S(x)\rangle_{\mathrm{np}} + 
P(\beta)\langle I \rangle \;,
\end{eqnarray}
where $Z(\beta)$ is a finite multiplicative  
renormalization of the discretized topological charge density~\cite{CDP88}, while 
$M(\beta)$ is an additive renormalization containing mixing to operators 
of equal or lower dimension and same quantum numbers, namely the action
density and the identity operator.
On the lattice $\chi^L(\beta)$ is measured during the Monte-Carlo simulation,
and  $\chi^{\rm cont}$ is extracted by subtracting the renormalizations,
which are computed non-perturbatively on the lattice by means of the 
``heating method''~\cite{DV92} ($Z$ and $P$ can be also computed 
perturbatively~\cite{FP93}).
The ``field theoretical method'' can be improved by using a smeared topological 
charge density 
operator~\cite{CDPV96}, built from the standard operator by replacing the 
fields $z$ and $\lambda_\mu$ with smeared fields $z^{\mathrm{smear}}$ and 
$\lambda_\mu^{\mathrm{smear}}$. 
In this way the renormalizations are strongly reduced and a much better 
accuracy is achieved for $\chi^{\rm cont}$.

\begin{figure}[t]
\begin{center}
\includegraphics[width=0.45\textwidth]{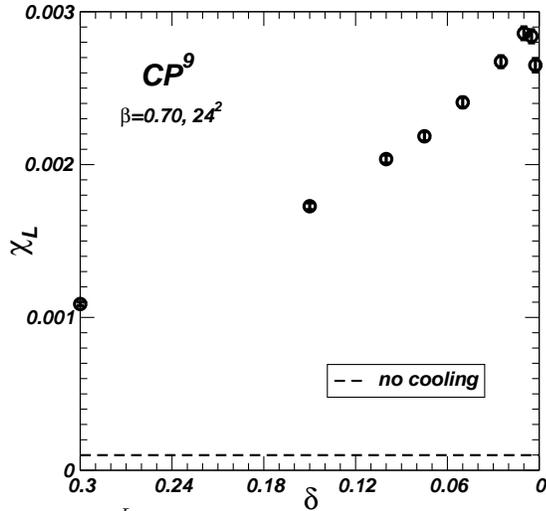}
\end{center}
\vspace{-1.54truecm}
\caption{$\chi^L$ as a function of $\delta$ at $\beta=0.70$ on a $24^2$ lattice
for CP$^9$.}
\label{chi_t_latt}
\end{figure}

\begin{figure}[t]
\begin{center}
\includegraphics[width=0.45\textwidth]{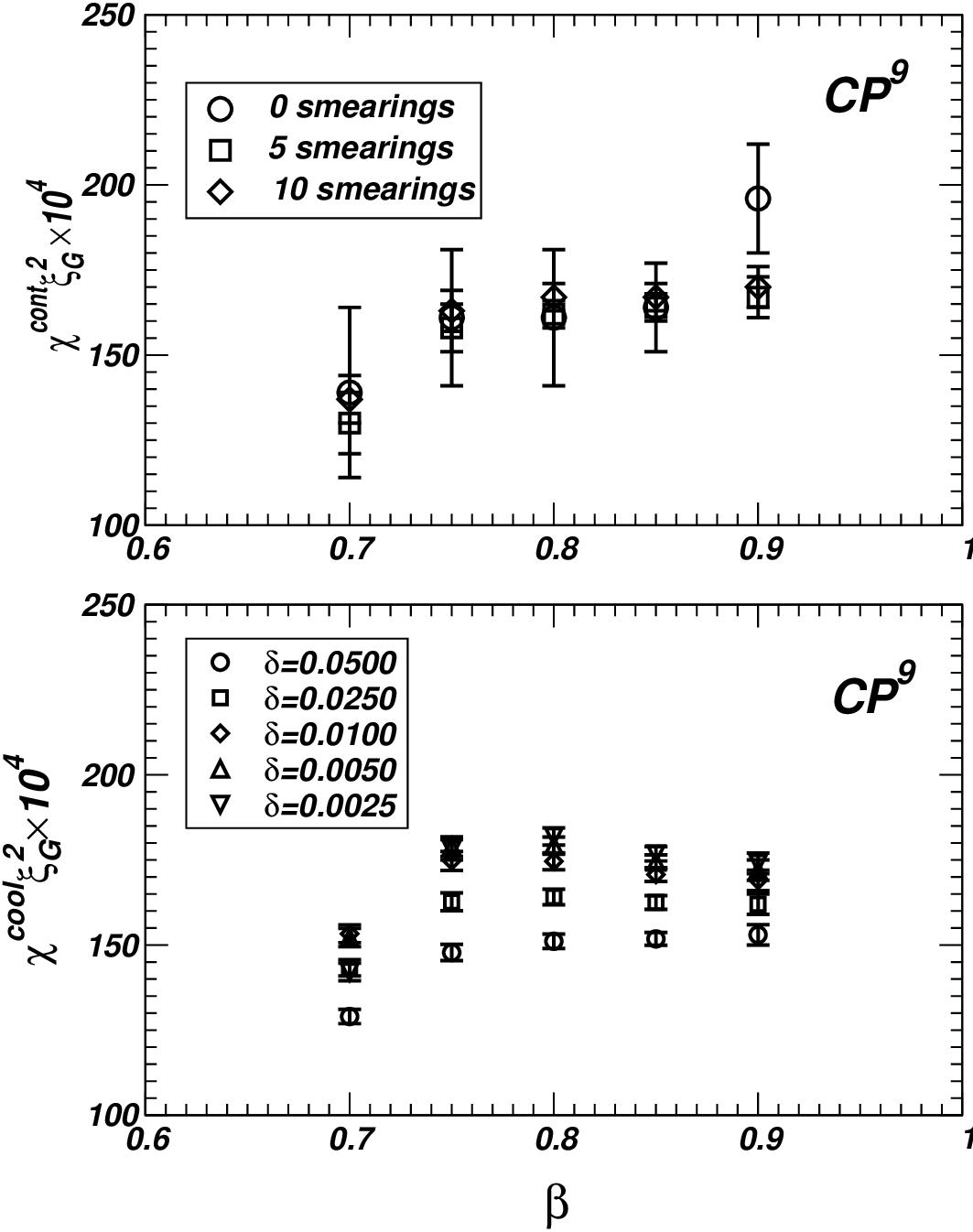}
\end{center}
\vspace{-1.54truecm}
\caption{$\chi^{\rm cont} \xi_G^2$ for CP$^9$ for several $\delta$ values
($L/\xi \simeq 10$ for all the points).}
\label{scaling}
\end{figure}

\begin{figure}[H]
\begin{center}
\includegraphics[width=0.45\textwidth]{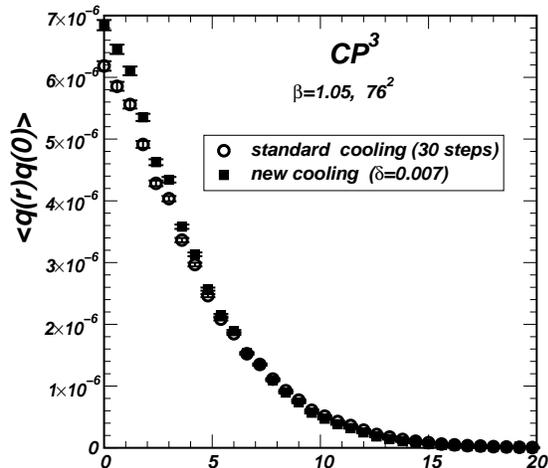}
\end{center}
\vspace{-1.54truecm}
\caption{``Shell'' correlation function of the 
lattice topological charge density ($r$ is the distance
between two lattice sites).}
\label{q_corr_1}
\end{figure}

We performed numerical simulations for $N=4$, $10$, $21$.
The simulation algorithm is a mixture of 4 microcanonical updates and 1 over-heat 
bath. For each simulation we collected 100K configurations after 
10K thermalization updating steps.
We used several values of the $\delta$ parameter in the ``new cooling''
algorithm, while for the ``Pisa cooling'' we considered $\delta_{\mathrm{Pisa}}
=0.2$.
To set the scale we have taken the correlation length $\xi_G$ defined 
as the second moment of the correlation function $\langle \mbox{Tr} P(x)\, P(0) 
\rangle$. 
In Fig.~\ref{chi_t_latt} we show the behavior of $\chi^L$
on configurations cooled by the ``new cooling'' with different 
$\delta$'s, while in Fig.~\ref{scaling} we plot $\chi^{\rm cont} \xi_G^2$ for the 
values of $\delta$ which correspond to the peak region in $\chi^L$ in 
Fig.~\ref{chi_t_latt} and compare the results with the field theoretical 
determination (with and without smearing). There is consistency
between the two determinations for all the considered values of $\delta$, which 
correspond to those for which the smoothing of the configuration is better.
We have also calibrated the three cooling techniques (number of cooling 
steps for the standard and Pisa cooling versus $\delta$ for the new cooling) in
order that the average energy on the ensemble cooled in the three different
ways is the same. Then, comparing by eye several thermal configurations obtained
after  equivalent amounts of the three coolings, we have observed that the distributions 
and the shape of the instanton bumps is roughly the same. Also the values of 
$\chi^{\rm cont} \xi_G^2$ obtained after ``equivalent'' coolings are in agreement
within the statistical errors. The only measurement where we could see a
discrepancy is that of the ``shell'' correlation function of the topological 
charge density $\langle q_L(r) q_L(0) \rangle$ (see Fig.~\ref{q_corr_1}), which
is slightly larger in the case of the new cooling method in the short distance
region, with respect to the standard and Pisa coolings.

Our conclusion is that, except for the last observation which deserves further
study, there is no appreciable difference between the three types of cooling we have
investigated.

\end{document}